\documentclass[pra,showpacs,preprintnumbers,amsmath,amssymb]{revtex4}
\usepackage[]{graphicx}
\usepackage[]{textcomp}

\newcommand{\be}{\begin{equation}}
\newcommand{\ee}{\end{equation}}
\newcommand{\bea}{\vspace{0.25cm}\begin{eqnarray}}
\newcommand{\eea}{\end{eqnarray}}

\def\PRL{{\it Phys. Rev. Lett.} }

\begin{document}
\title{ Two-mode squeezed vacuum and squeezed light in correlated interferometry}

\author{I.~Ruo Berchera$^1$, I.~P.~Degiovanni$^1$, S.~Olivares $^3$, N. Samantaray$^{1,2}$, P.Traina $^1$, and M.~Genovese$^1$ }

\affiliation{$^1$INRIM, Strada delle Cacce 91, I-10135 Torino, Italy}

\affiliation{$^2$ Politecnico di Torino }

\affiliation{$^3$ Dipartimento di Fisica, Universit\`a degli Studi di Milano, Via Celoria 16, I-20133 Milano, Italy}

\begin{abstract}
We study in detail a system of two interferometers aimed to the detection of extremely faint phase-fluctuations. This system can represent a breakthrough for detecting a faint correlated signal that would remain otherwise undetectable even using the most sensitive individual interferometric devices, that are limited by the shot noise. If the two interferometers experience identical phase-fluctuations, like the ones introduced by the so called ``holographic noise'', this signal should emerge if their output signals are correlated, while the fluctuations due to shot noise and other independent contributions will vanish.
We show how the injecting quantum light in the free ports of the interferometers can reduce the photon noise of the system beyond the shot-noise, enhancing the resolution in the phase-correlation estimation. We analyze both the use of two-mode squeezed vacuum or twin-beam state (TWB) and of two independent squeezing states. Our results basically confirms the benefit of using squeezed beams together with strong coherent beams in interferometry, even in this correlated case. However, mainly we concentrate on the possible use of TWB, discovering interesting and probably unexplored areas of application of bipartite entanglement and in particular the possibility of reaching in principle surprising uncertainty reduction.
\end{abstract}
\maketitle
\section{Introduction}

The possibility of increasing the performances of interferometers by using quantum light represents one of the most interesting use of quantum states for overcoming classical limits of measurements.
The first approach proposed to this aim is based on exploiting squeezed light for reducing the noise level in interferometers \cite{cav} and found recently application in gravitational waves detectors \cite{mc,ab}.
A second approach is considering the use of entanglement in phase estimation and, in particular, the possibility offered by the use of NOON states \cite{noon}. However, even if this approach presents a significant conceptual interest and could find very interesting applications in the future, nowadays the difficulty in producing high $N$ entangled states and the fragility to noise and losses of these schemes strongly limits their real possible use.
More recently, correlation in photon number in two-mode squeezed vacuum or twin-beam state (TWB) \cite{Chekhova2015} has been demonstrated to be an important tool  for beating shot noise \cite{n1} and for realising a first quantum protocol effectively robust against  noise and losses \cite{n2}.
These results prompted to study the possibility of improving the so called ``holometer'' by exploiting quantum light, and in particular squeezed one or TWBs \cite{prl}.
The holometer is a double Michelson Interferometer (MI) addressed to detect the so called ``holographic noise'' (HN), namely a basic form of noise conjectured in quantum gravity theories that would derive from a non-commutativity of the spatial degrees of freedom at the Planck scale \cite{hog}. This noise, albeit very small, should be correlated when the two MIs are parallel, such to be in the respective light cones, while should be uncorrelated when one arm is rotated to be oriented in the opposite direction for the two MIs.
The evident huge impact that the discovery of holographic noise, the first eventual evidence of quantum gravity effects, would present, \cite{qg} motivates an accurate analysis of the possibility of improving the holometer performances.
In this paper we detail and complete the analysis of Ref.~\cite{prl} identifying operative situations where the use of quantum light would allow to greatly increase the performances of a double interferometer like the holometer.

More in detail, in \cite{prl}  we investigated an unusual but potentially powerful system consisting of two interferometers whose correlation of output ports signals is measured (see Fig.\ref{scheme}). This kind of double interferometric system can represent a breakthrough for detecting a faint correlated signal that would remain otherwise hidden even using the most sensitive individual interferometric devices, limited by the shot noise. On the other side if the two interferometers are in the experience identical fluctuation, this signal should emerge in a correlation measurement of their output, while the fluctuations due to shot noise and other independent contributions will vanish.

The first experimental realization of this scheme using coherent beams (stabilized lasers), exactly for HN detection, is already being implemented at Fermilab. Other applications can be envisaged such as for new generation of gravitational wave detectors.

In \cite{prl} we have introduced a rigorous quantum model for describing the system. As opposed to standard phase measurement in a single interferometer, which involved first order expectation value of the output, in the double interferometric scheme the quantity under estimation is the covariance of the two outputs, which is a second order quantity, thus the associated uncertainty is a fourth order function. Notwithstanding this difference, we demonstrated how the injection of quantum light at the input ports, which would remain unused in classical Holometer configuration, can boost the sensitivity of the device. In addition to the classical intense coherent beam, we considered both the use of independent squeezed beam (SQB) and correlated state such as the TWB. The ideal experiment described theoretically there, is however not suitable for a practical experimental implementation. In particular, the setting of the central phases the two interferometers exactly at $\phi_{1,0}=\phi_{2,0}=0$ which provides the optimal quantum enhancement, is indeed critical, because minimal deviations from this working regime completely compromise the advantages of the quantum strategy. Furthermore, the balanced readout configuration explored in SQB case would require simultaneously high dynamic range, fast and high resolution detectors that are not yet available.

Here we present a framework in which a more complete and general study of the double interferometric system is provided, leading also to the depiction of a more experiment-oriented configuration of the system in terms of readout strategy and parameters choice.

In Sec. \ref{The interferometric scheme} we present the description of the scheme. In Sec. \ref{Correlations at the read-out ports} we analyse the correlation properties at the output ports of the interferometer in the case of TWB,  demonstrating that for a proper choice of the relative phase $\psi$ between the coherent beam and the TWB, the energy transfer from the coherent beam due to the mixing allows the generation of either bright quantum correlation ($\psi=\pi/2$) or anti-correlation ($\psi=0$) in the photon number, as witnessed from the Noise Reduction Factor (NRF) value below unity. Furthermore, we demonstrate that there are two regimes with different behaviour of the system: (A), in which a strict choice of the central phases of the interferometers makes the contribution to the coherent light at the dark port negligible, i.e. only quantum light is detected. In principle it leads to the optimal correlation (only bounded by losses), but turns out quite challenging in practice, since would require extremely precise stabilization of the interferometers, especially if the coherent beam is intense.(B), in which the major component of the signal at the output ports is due to the coherent beam contribution. This is a more common and realistic working condition. There is a sudden transition between these two regimes.

In Sec. \ref{Estimation of phase-correlation (holographic noise)} we describe in detail a model establishing the connection between a generic measurement operator (observable) and the estimation of the phase-covariance introduced by a correlated faint phase signal such as HN. Then, we focus on two specific quantum strategies: either on the use of TWB state and the measure of the photons number difference, of the use of two independent squeezed states and the measure of quadratures covariance. In both cases we evaluate the lower bound to the uncertainty in the phase-covariance-estimation given by photon noise, in function of the fundamental parameters: the interferometers central phases $\phi_{i,0}$ ($i=1,2$), the quantum and classical beam intensities $\lambda$ and detection efficiency $\eta$. The Results are reported in Sec. \ref{Results}. For rather challenging conditions, namely almost ideal efficiency and perfect control of the stability of the interferometers central phases (regime (A)), TWB could deliver extraordinary advantage due to its photon number correlation at the quantum level (entanglement). This regime corresponds to the situation analyzed in \cite{prl}. Conversely, there exists a less demanding regime ((B)-regime described before), in which quantum strategies provide good enhancement in a more favorable experimental condition. In this case,both for TWB and SQB the expression of the minimal uncertainty presents the usual scaling with losses, ($\propto1-\eta$), and with the quantum light intensity, ($\propto1/\lambda$), typical of single phase estimation using strong local oscillator and squeezed light.

Finally we draw the conclusions in Sec. \ref{conclusions}

\section{The interferometric scheme}\label{The interferometric scheme}

\begin{figure}[htb]
\centering
\includegraphics[width=14cm]{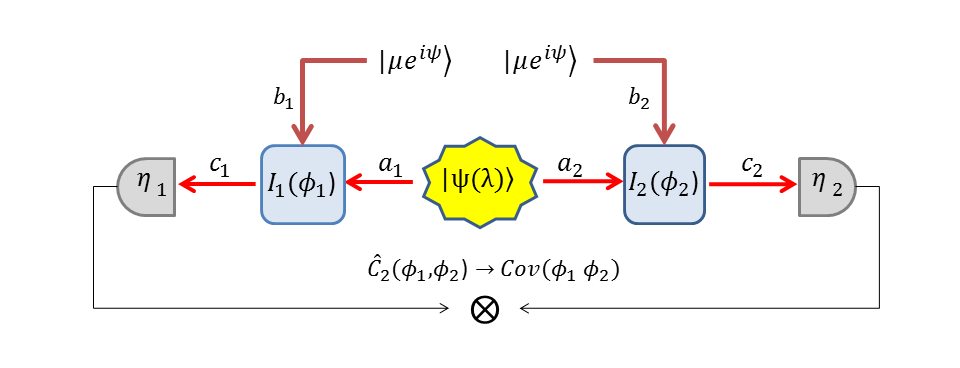} \caption{Ideal scheme of the proposed experiment. The two modes of a twin beam are mixed with two identical coherent states at the input beam-splitters of two coupled interferometers (holometer). The covariance of the phase noise at the outputs is studied in order to detect very faint (but correlated) noise.}
\label{scheme}
\end{figure}

Let us consider a system as depicted in Fig.\ref{scheme}. Two interferometers $\mathcal{I}_{i}$ ($i=1,2$) are injected at the ports denoted by the mode annihilation operators $b_{i}$ by a couple of identical coherent beams $|\sqrt{\mu} e^{i\psi}\rangle_{b_{i}}$, while the remaining ports identified by the mode operator $a_{i}$ (unused in the classical scheme) are fed with a quantum state $\vert\mathrm{\Psi}(\lambda)\rangle_{a_1,a_2}$, where $\lambda$ is the mean number of photon in each mode. The readout ports are denoted by the mode operator $c_{i}$ which will be function of the phases shifts $\phi_{i}$ among the arm of each interferometer, $c_{i}=c_{i}(\phi_{i})$. Therefore, a final combination of the outputs results in an observable $\widehat{C}\left(c_{1}, c_{2}, h.c.\right)=\widehat{C}\left(\phi_{1},\phi_{2}\right)$. A proper choice of the operator $\widehat{C}$ leads to an estimation of the phase-noise correlation. Here, it is useful to recall the properties that the input-output operator relations of a linear interferometer (for example a Michelson-type) are equivalent to the ones of a beam splitter (BS) with transmission coefficient $\tau=\cos^{2}(\phi/2)$.

The losses in the system are taken into account by considering in both channels two identical detectors with quantum efficiency $\eta$, formally re-defining the output operators with the substitution $c_{i}\rightarrow \sqrt{\eta}c_{i}$ in the normal-ordering products. For example the photon number operator will be $N_{i}\equiv \eta c_{i}^{\dag}c_{i}$.

If only the classical field is injected, the photon counting statistics at the output ports is simply the one of coherent beams after reflection probability $(1-\tau_{i})$ and detection probability $\eta$,
\begin{eqnarray} \label{COHstat}
  \langle N\rangle^{coh}_{\eta\tau_{i}} &=& \eta(1-\tau_{i})\mu,\\\nonumber
  \langle\delta N^{2}\rangle^{coh}_{\eta\tau_{i}} &=& \langle N_{i}\rangle^{coh}_{\eta\tau_{i}}=\eta \mu (1-\tau_{i}), \\
  \langle\delta N_{1} \delta N_{2}\rangle^{coh}_{\eta\tau_{1}\tau_{2}} &\equiv& 0. \nonumber
 \end{eqnarray}

Now we consider two possible quantum states feeding the free input ports of $\mathcal{I}_{i}$ and two related readout strategies.

{\it Readout strategy 1: TWB} -- The TWB correlated state can be expressed in the Fock bases $\left\{\vert m \rangle_{a_{i}}\right\}$ as

\begin{equation}\label{TWB}
 \vert\mathrm{\Psi}(\lambda)\rangle_{a_{1},a_{2}}=\frac{1}{\sqrt{1+\lambda}}\sum_{m=0}^{\infty}\left(e^{i\theta}\sqrt{\frac{\lambda}{1+\lambda}}\right)^{m}\vert m,m\rangle_{a_{1},a_{2}}
\end{equation}
where $\theta$ is the global phase, which we set in the following to $\theta=0$ without loss of generality.
The TWB presents perfect correlations in the photon number $m_{i}\equiv a_{i}^{\dag}a_{i}$ meaning that ${}_{a_1,a_2}\langle\langle{\rm TWB}|(\hat{N}_{a_1}-\hat{N}_{a_2})^{M}|{\rm
TWB}\rangle\rangle_{a_1,a_2}=0$, $\forall M>0$. It implies for example that variance of the photons numbers difference $\langle\delta(m_{1}- m_{2})^{2}\rangle$ (with $\delta m\equiv m-\langle m\rangle$) is identically null if losses are neglected. It also suggests to choose the measurement operator in the same form  $\widehat{C}\left(\phi_{1},\phi_{2}\right) =\langle(N_{1}- N_{2})^{M}\rangle$, since this should correspond to a reduction of the photon noise in the measurement, finally improving the sensitivity.

By using the equivalence between interferometers and BSs mentioned before, we can calculate the photon statistics of TWB transmitted to the output ports (in absence of classical coherent field) and detected with quantum efficiency $\eta$. The mean photon number, the variance and the covariance are respectively:
\begin{eqnarray} \label{TWBstat}
  \langle N\rangle^{TWB}_{\eta\tau_{i}} &=& \eta \tau_{i}\lambda \\\nonumber
  \langle\delta N^{2}\rangle^{TWB}_{\eta\tau_{i}} &=& \eta \tau_{i} \lambda (1+\eta \tau_{i} \lambda) \\
  \langle\delta N_{1} \delta N_{2}\rangle^{TWB}_{\eta\tau_{1}\tau_{2}} &=& \eta^{2}\tau_{1}\tau_{2}\lambda(1+\lambda)\nonumber
 \end{eqnarray}

{\it Readout strategy 2: two squeezed states} --  The product of two single mode squeezed vacuum states writes:
 $$|\xi \rangle_{a_1}\otimes|\xi \rangle_{a_2} = S_{a_1}(\xi)S_{a_2}(\xi)|0\rangle_{a_1}\otimes |0\rangle_{a_2}$$
where $S_{a_i}(\xi) = \exp[ \xi~(a_i^\dag)^2 - \xi^*~(a_i)^2]$ is the squeezing operator. If we set $\xi=|\xi| e^{i \theta_\xi}$,  then $\lambda=\sinh^2 |\xi|$ represents the average number of photons of the squeezed vacuum, taken equal in both the modes.

Defining the the quadrature of the field as $x_{i}\equiv \frac{a_{i}+a_{i}^{\dagger}}{\sqrt{2}}$ and $y_{i}\equiv \frac{a_{i}-a_{i}^{\dagger}}{i\sqrt{2}}$, and supposing $y_{i}$ the squeezed and $x_{i}$ the anti-squeezed one, it is known that in the single interferometer the injection of the squeezed field provides a fixed factor $\langle\delta y_{i}^{2}\rangle=e^{-2|\xi|}$ of resolution enhancement for arbitrary brightness of the coherent beam \cite{cav}. It is expected that the increased resolution in the estimation of the phase shifts $\phi_{1}$ and  $\phi_{2}$ separately reflects in a better estimation of their correlation if the correlation of the squeezed quadrature $X_{i}$ of the output modes $c_{i}$ are considered, namely $\widehat{C}=X_{1}\cdot X_{2}$.

\section{Correlations at the readout ports}\label{Correlations at the read-out ports}

As a figure of merit for the correlations at the read-out ports, we study the noise reduction parameter $NRF_{\pm}\equiv\langle\delta(N_{1}\pm N_{2})^{2}\rangle/\langle N_{1}+N_{2}\rangle$ \cite{mas,Lamperti2014,m,Lanz2008}. It represents the ratio between the variance of the photon number sum (difference) and the corresponding shot noise limit. The $NRF_{-}<1$ it is a well known condition of non-classicality for the correlations of a bipartite state and its value also determines the quantum enhancement achievable in certain sensing and imaging protocols \cite{n1}.
For the same reason, $NRF_{+}<1$ could be be interpreted as strong signature of anti-correlation of the photon number beyond classical limits.
The BS-like transformation allows  evaluating the fluctuation of the fields at the output ports in function of the input field. In particular one gets

\begin{eqnarray}\label{Ni}
\langle N_{i}\rangle&=&  \langle N\rangle^{TWB}_{\eta\tau_{i}}+\langle N\rangle^{coh}_{\eta\tau_{i}}
\end{eqnarray}
\begin{eqnarray}\label{V(Ni)}
\langle\delta N_{i}^{2}\rangle&=& \langle\delta N^{2}\rangle^{TWB}_{\eta\tau_{i}}+\langle\delta N^{2}\rangle^{coh}_{\eta\tau_{i}} + 2 \langle N\rangle^{TWB}_{\eta\tau_{i}} \langle N\rangle^{coh}_{\eta\tau_{i}}
\end{eqnarray}
\begin{eqnarray}\label{cov(N1N2)}
\langle\delta N_{1} \delta N_{2}\rangle &=& \langle\delta N_{1} \delta N_{2}\rangle^{TWB}_{\eta\tau_{1}\tau_{2}}- 2 \sqrt{\langle\delta N_{1} \delta N_{2}\rangle^{TWB}_{\eta\tau_{1}\tau_{2}} \langle N\rangle^{coh}_{\eta\tau_{1}} \langle N\rangle^{coh}_{\eta\tau_{2}}}\cos[2 \psi ]
\end{eqnarray}
and the explicit expression, function of the parameters, can be directly obtained by substituting the quantities according to Eq.s (\ref{COHstat}) and (\ref{TWBstat}).
In particular Eq. (\ref{cov(N1N2)}) shows that the covariance is composed by the TWB covariance $\langle\delta N_{1} \delta N_{2}\rangle^{TWB}_{\eta\tau_{1}\tau_{2}}$  and a second term containing the phase of the coherent field $\psi$, originated by the BS interaction of the two fields. Interestingly, the choice of $\psi=\pi/2$ maximize the covariance, while for $\psi=0$ the covariance can even be negative (anti-correlation of photon numbers).

The $NRF_{\pm}$ can be easily calculated from Eq.s (\ref{Ni},\ref{V(Ni)},\ref{cov(N1N2)}) by exploiting the identity $\langle\delta(N_{1}\pm N_{2})^{2}\rangle=\langle\delta N_{1}^{2}\rangle+\langle\delta N_{2}^{2}\rangle\pm 2 \langle\delta N_{1}\delta N_{2}\rangle$.  Using the same notation we have

\begin{equation}\label{NRF}
NRF_{\pm}=\frac{\langle\delta(N_{1}\pm N_{2})^{2}\rangle^{TWB}_{\eta\tau}+ 2 \langle N\rangle^{coh}_{\eta\tau} \left( 1 + 2 \langle N\rangle^{TWB}_{\eta\tau} \pm 2\sqrt{ \langle\delta N_{1} \delta N_{2}\rangle^{TWB}_{\eta\tau}}\cos[2 \psi ]\right)}{2 \langle N\rangle^{coh}_{\eta\tau}+2 \langle N\rangle^{TWB}_{\eta\tau} }
\end{equation}

For simplicity, in Eq. (\ref{NRF}) and in the following we have assume $\tau_{1}=\tau_{2}=\tau$. We note that in general the $NRF_{-}$ is minimized for $\psi=\pi/2$ (corresponding to the optimization of the photon number correlation), while $NRF_{+}$ is minimized when $\psi=0 $ (corresponding to the optimization of the anti-correlation). The $NRF_{-}(\psi=\pi/2)$ and $NRF_{+}(\psi=0)$ are plotted in Fig.\ref{NRFplot}. In order to analyze the behaviour shown in the figures, and for the forthcoming discussion of the results concerning the  phase-covariance estimation in Sec. \ref{Results}, it is useful to distinguish two regimes:

\begin{figure}[htb]
\centering
\includegraphics[width=9cm]{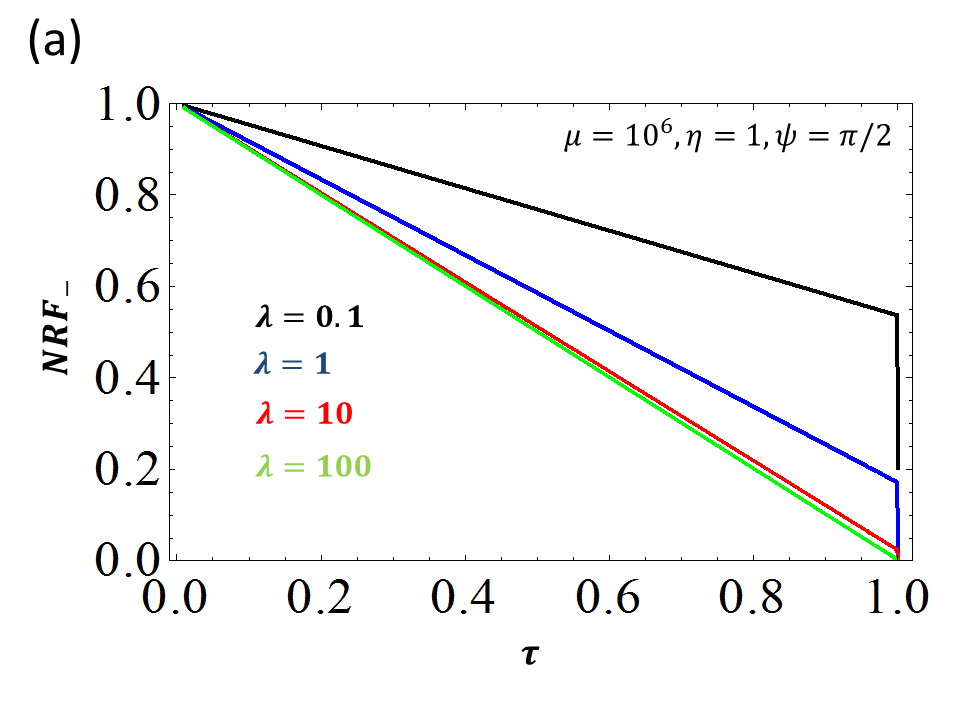}
\includegraphics[width=9cm]{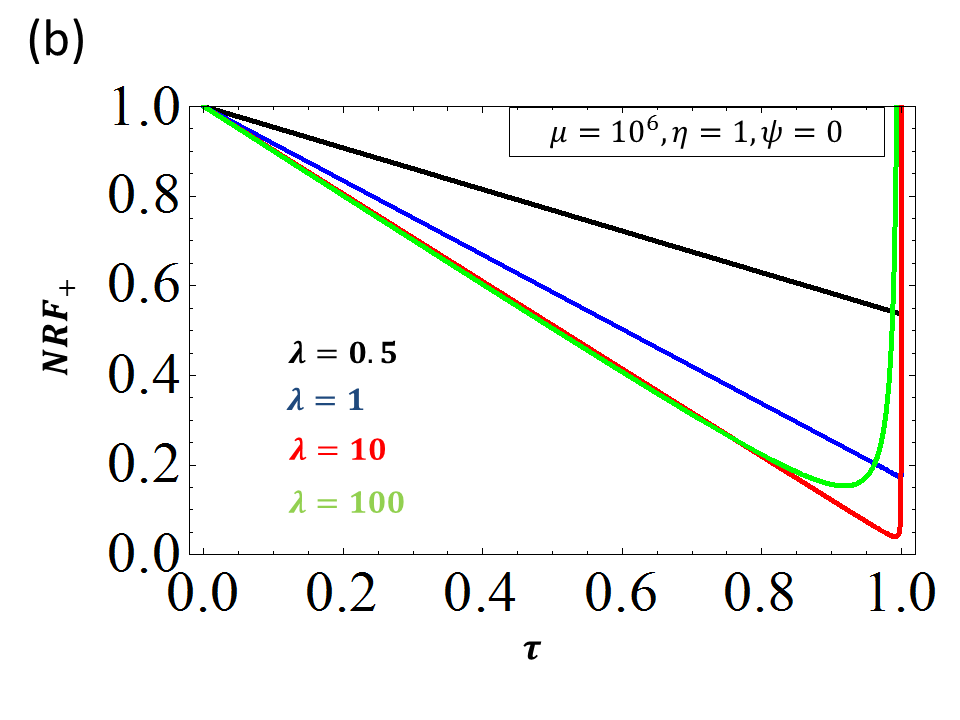}
\caption{Noise Reduction Factor versus the transmission coefficient $\tau$ of the TWB modes at the read-out ports of the interferometers.  Different colored-lines stand for different mean number of photons of the TWB ($\lambda$): a) is the noise reduction factor ($NRF_{-}$) exploiting quantum correlation at the output ports, for $\psi=\Pi/2$, while b) is the noise reduction factor ($NRF_{+}$) exploiting quantum anti-correlation, for $\psi=0$. Here we consider the ideal lossless case, $\eta=1$, and we set $\mu=10^{6}$.}
\label{NRFplot}
\end{figure}

\subsection*{Regime-(A): TWB-like correlations}

Referring to Eq.(\ref{NRF}), when the intensity at the read-out port is dominated by the TWB, i.e. $\langle N\rangle^{coh}_{\eta\tau}\ll \langle N\rangle^{TWB}_{\eta\tau}$ or explicitly $k\equiv\mu\left(1-\tau\right)/\tau\lambda \ll1$, the noise reduction factor reduces to the one of TWB, $NRF_{\pm}\simeq\langle\delta(N_{1}\pm N_{2})^{2}\rangle^{TWB}_{\eta\tau}/2 \langle N\rangle^{TWB}_{\eta\tau}$:
\begin{equation}
NRF_{-}\simeq (1-\eta\tau )+\eta\tau (1+2\lambda -2\sqrt{\lambda(1+\lambda)})\kappa\simeq(1-\eta  \tau).
\label{NRF-(A)}
\end{equation}
\begin{equation}
NRF_{+}\simeq 1+\eta\tau (2\lambda +1)
\label{NRF+(A)}
\end{equation}
recovering the expression of noise reduction factor for TWB in presence of losses \cite{n1}. Of course the condition $\langle N\rangle^{coh}_{\eta\tau}\ll \langle N\rangle^{TWB}_{\eta\tau}$ appears quite challenging to achieve in the relevant case of practical interest in which the coherent mode is largely populated. Larger is $\mu$, closer to unity have to be the equivalent-transitivity $\tau$ of the interferometers. In Fig. \ref{NRFplot} it corresponds to the region $\tau\sim1$, where $NRF_{-}$ drops to zero, clearly behaving as a singular point. We observe that in the same regime, as indicated by Eq. (\ref{NRF+(A)}), the $NRF_{+}$ (marking the anti-correlation) rapidly grows accordingly. This regime has been studied in \cite{prl}. Even if in principle it allows  exploiting the perfect TWB-like correlation, with a large classical power circulating into the interferometer, thus obtaining surprising quantum enhancement in the phase correlation estimation, the fragility of this regime has demanded for a more extended exploration of the parameter space. This is one of the motivation of this extended work.

\subsection*{Regime-(B): bright quantum correlation}
When the coherent power reflected to the measuring port is much higher than the transmitted power of TWB $\langle N\rangle^{coh}_{\eta\tau}\gg \langle N\rangle^{TWB}_{\eta\tau}$, i.e $\mu\left(1-\tau\right)>> \lambda\tau$, Eq.(\ref{NRF}) reduces to
$$NRF_{\pm}\simeq1 + 2 \langle N\rangle^{TWB}_{\eta\tau} \pm 2\sqrt{ \langle\delta N_{1} \delta N_{2}\rangle^{TWB}_{\eta\tau}}\cos[2 \psi ].$$
Introducing the explicit expressions of the various moments of the photon number distribution distributions we have
\begin{equation}
NRF_{-}(\psi=\pi/2)=NRF_{+}(\psi=0)\simeq1+2\eta\tau\left( \lambda- \sqrt{\lambda(1+\lambda)}\right)\simeq 1-\tau\eta+ \frac{\eta\tau}{4\lambda},
\label{NRF+-(B)}
\end{equation}
where in the last equality we have also considered the limit of high intensity TWB, i.e.  $\lambda\gg1$. It is worth to be notice that the $NRF$, for the proper choice of the phase of the classical fields, is always smaller than 1, whatever the intensity of TWB and losses. Thus, $N_{1}$ and $N_{2}$ are always correlated (or anti-correlated) beyond the classical limit. It is possible to switch between quantum correlation and quantum anti-correlation just by acting on the phase $\psi$ of the classical fields. Even more interesting, the correlation can be extremely bright, because the mean number of photon at the read-out ports is determined by the brightness of the classical beam $\langle N\rangle^{coh}_{\eta\tau}= \eta(1-\tau)\mu$, which can be increased almost arbitrarily in real experiments. It is clear from Eq. (\ref{NRF+-(B)}) that the highest correlation is obtained when $\lambda\gg1$ and at the same time $\tau\sim1$. For example in Fig. \ref{NRFplot} for plausible values of $\tau\sim0.90$ and $\lambda=10 $ the NRF is 0.1, and the mean intensity of the output signal is  $(1-\tau)\mu=10^{5}$ photons.


\section{Estimation of phase correlation (holographic noise)}
\label{Estimation of phase-correlation (holographic noise)}
Since we the phase fluctuations due to the holographic noise
are expected to be extremely small, we can expand $\widehat{C}(\phi_1, \phi_2)$ around the chosen central
values $\phi_{1,0}$, $\phi_{2,0}$, namely:
\begin{eqnarray}
\widehat{C}(\phi_1, \phi_2)&=&\widehat{C}(\phi_{1,0}, \phi_{2,0})+ \Sigma_i ~ \partial_{\phi_{i}}
\widehat{C}(\phi_{1,0}, \phi_{2,0})~ \delta \phi_i + \frac{1}{2} \Sigma_i ~
\partial_{\phi_{i},\phi_{i}}^{2} \widehat{C}(\phi_{1,0}, \phi_{2,0})~ \delta \phi_{i}^2 \nonumber \\
&\hbox{}&+
\partial_{\phi_{1},\phi_{2}}^{2} \widehat{C}(\phi_{1,0}, \phi_{2,0}) ~ \delta \phi_{1} \delta
\phi_{2}+ \mathcal{O}(\delta \phi^3)
\label{C}
\end{eqnarray}
where  $\delta \phi_{i}= \phi_{i}-\phi_{i,0}$, and
$\partial_{\phi_{1}^h, \phi_{2}^k}^{h+k}\widehat{C}(\phi_{1,0},
\phi_{2,0}) $ is the $(h+k)$-th order derivative of
$\widehat{C}(\phi_{1}, \phi_{2}) $ calculated  at $\phi_{i}=\phi_{i,0}$, $i,j=1,2$
\par
In order to reveal the HN, the holometer exploits two different
configurations: the one, ``$\parallel$", where HN correlates the
interferometers, the other, ``$\perp$ ", where the effect of HN
vanishes.  The statistical properties of the phase-shift (PS)
fluctuations due to the HN may be described by the joint probability
density functions $f_\parallel (\phi_{1}, \phi_{2})$ and $f_\perp
(\phi_{1}, \phi_{2})$. We make two reasonable hypotheses about
$f_x (\phi_{1}, \phi_{2})$, $x=\parallel,\perp$.  First, the marginals
$\mathcal{F}^{(i)}_{x} (\phi_{i})=\int \mathrm{d} \phi_{j } f_x
(\phi_{i}, \phi_{j})$, $i,j=1,2$ with $i\neq j$, are exactly the same
in the two configurations, i.e.
$\mathcal{F}^{(i)}_{\parallel}(\phi_{i})=\mathcal{F}^{(i)}_{\perp}(\phi_{i})$:
one cannot distinguish between the two configurations just by
addressing one interferometer. Second, only in configuration
``$\perp$'' it is $f_\perp (\phi_{1}, \phi_{2})=
\mathcal{F}^{(1)}_{\perp}(\phi_{1})\mathcal{F}^{(2)}_{\perp} (\phi_{2})$,
i.e., there is no correlation between the PSs due to the HN.  Now, the
expectation of any operator $\widehat{O}(\phi_1,\phi_2)$ should be
averaged over $f_x$, namely, $\langle\widehat{O}(\phi_1,\phi_2)\rangle
\to \mathcal{E}_{x}\left[\widehat{O}(\phi_1,\phi_2)\right] \equiv \int
\langle\widehat{O}(\phi_1,\phi_2)\rangle ~ f_x (\phi_{1},
\phi_{2}) ~ \mathrm{d} \phi_{1} ~ \mathrm{d} \phi_{2}$.  In
turn, by averaging the expectation of Eq.~(\ref{C}), we have:
\begin{eqnarray}
\mathcal{E}_x \left[ \widehat{C}(\phi_1, \phi_2) \right]
&=&
\langle \widehat{C}(\phi_{1,0}, \phi_{2,0}) \rangle
+  \frac{1}{2} \Sigma_i ~
\langle \partial_{\phi_{i},\phi_{i}}^{2}  \widehat{C}(\phi_{1,0},
\phi_{2,0}) \rangle ~ \mathcal{E}_x \left[\delta \phi_{i}^2\right]
\nonumber \\
& \hbox{} & +
 \langle \partial_{\phi_{1},\phi_{2}}^{2}\widehat{C}(\phi_{1,0},
\phi_{2,0}) \rangle~  \mathcal{E}_x \left[ \delta \phi_{1} \delta
\phi_{2}\right]+ \mathcal{O}(\delta \phi^3)
 \label{EC}
\end{eqnarray}
where we used $\mathcal{E}_x \left[\delta
  \phi_k\right]=0$. Then, according to the assumption on $f_x (\phi_{1}, \phi_{2})$ we have $\mathcal{E}_\parallel \left[\delta
  \phi_{k}^2\right]= \mathcal{E}_\perp \left[\delta
  \phi_{k}^2\right]$ and $\mathcal{E}_\perp \left[\delta \phi_{1} \delta
\phi_{2}\right] =\mathcal{E}_\perp \left[\delta \phi_{1} \right] \mathcal{E}_\perp \left[\delta
\phi_{2}\right]= 0$, and from Eq.~(\ref{EC}) follows that the phase-covariance may be written as:
\begin{equation}\label{PSs-Cov}
  \mathcal{E}_\parallel \left[ \delta \phi_{1} \delta \phi_{2} \right]
  \approx
  \frac{\mathcal{E}_\parallel \left[ \widehat{C}(\phi_1, \phi_2)
      \right]-\mathcal{E}_\perp \left[ \widehat{C}(\phi_1, \phi_2)\right]}{\langle \partial_{\phi_{1},\phi_{2}}^{2}\widehat{C}(\phi_{1,0},\phi_{2,0}) \rangle },
\end{equation}
that is proportional to the difference between the mean values of the operator
$\widehat{C}(\phi_1, \phi_2)$ as measured in the two configurations
``$\parallel$ and ``$\perp$''.

Indeed, one has to reduce as much as possible the uncertainty associated with its measurement:
\begin{equation}\label{U}
 \mathcal{U}(\delta \phi_{1} \delta \phi_{2}) \approx
\sqrt{\frac{\mathrm{Var}_\parallel \left[ \widehat{C}(\phi_1,\phi_2)
 \right] + \mathrm{Var}_\perp  \left[ \widehat{C}(\phi_1,\phi_2) \right]}{\left[ \langle
\partial_{\phi_{1},\phi_{2}}^{2}  \widehat{C}(\phi_{1,0}, \phi_{2,0}) \rangle \right]^2 }}, \quad
(\delta \phi_{1},\delta\phi_{2} \ll 1)
\end{equation}
where $\mathrm{Var}_x \left[ \widehat{C}(\phi_1,\phi_2) \right] \equiv
\mathcal{E}_x \left[ \widehat{C}^2 (\phi_1,\phi_2) \right] -
\mathcal{E}_x \left[ \widehat{C}(\phi_1,\phi_2) \right]^2$.

Under the same hypotheses used for deriving Eq. (\ref{PSs-Cov}) we can calculate the variance of $\widehat{C}(\phi_1, \phi_2)$ as
\begin{eqnarray}\label{VarC}
\mathrm{Var}_x \left[ \widehat{C}(\phi_1,\phi_2) \right]=\mathrm{Var} \left[ \widehat{C}(\phi_{1,0},\phi_{2,0}) \right]+ \Sigma_k ~ A_{kk}~\mathcal{E}_x \left[\delta \phi_{k}^2\right]+ A_{12}~\mathcal{E}_x \left[\delta \phi_{1}\delta \phi_{2}\right]+\mathcal{O}(\delta \phi^3)
\end{eqnarray}
where:
\begin{eqnarray}\label{Akj}
A_{kk}&= &\langle \widehat{C}(\phi_{1,0},\phi_{2,0}) \partial_{\phi_{k},\phi_{k}}^{2}\widehat{C}(\phi_{1,0},\phi_{2,0})\rangle\\\nonumber
&+&\langle [\partial_{\phi_{k}} \widehat{C}(\phi_{1,0},\phi_{2,0})]^{2} \rangle-\langle \widehat{C}(\phi_{1,0},\phi_{2,0})\rangle\langle  \partial_{\phi_{k},\phi_{k}}^{2}\widehat{C}(\phi_{1,0},\phi_{2,0})\rangle\\\nonumber
A_{12}&=&2\langle \widehat{C}(\phi_{1,0},\phi_{2,0}) \partial_{\phi_{1},\phi_{2}}^{2}\widehat{C}(\phi_{1,0},\phi_{2,0})\rangle\\\nonumber
&+&2\langle \partial_{\phi_{1}} \widehat{C}(\phi_{1,0},\phi_{2,0})\partial_{\phi_{2}} \widehat{C}(\phi_{1,0},\phi_{2,0}) \rangle-\langle \widehat{C}(\phi_{1,0},\phi_{2,0})\rangle\langle  \partial_{\phi_{1},\phi_{2}}^{2}\widehat{C}(\phi_{1,0},\phi_{2,0})\rangle
\end{eqnarray}
Analyzing expression (\ref{VarC}), we note the presence of a
zeroth-order contribution that does not depend on the PSs intrinsic
fluctuations, and represents the quantum photon noise of the
measurement described by the operator $\widehat{C}(\phi_1,\phi_2)$
evaluated on the optical quantum states sent into the holometer. The
statistical characteristics of the phase noise enter as second-order
contributions in Eq.~(\ref{VarC}) from each interferometer plus a
contribution coming from phase correlation between them.

This work addresses specifically the problem of reducing the photon
noise below the shot noise in the measurement of the HN, therefore in
the following, we will assume
the zero-order contribution being the dominant one. Of course, this
means to look for the HN in a region of the noise spectrum that is
shot-noise limited. Since the HN is expected up to frequencies of tens
MHz, it follows that all the sources of mechanical vibration noise are
suppressed. Therefore, the zero-order uncertainty that we will study here is
\begin{equation} \label{U0}
\mathcal{U}^{(0)} = \frac{\sqrt{2\, \mathrm{Var}\left[ \widehat{C}(\phi_{1,0 }, \phi_{2,0}) \right]}}{\left|
 \langle \partial_{\phi_{1},\phi_{2}}^{2} \widehat{C}(\phi_{1,0},
 \phi_{2,0}) \rangle \right|},
\end{equation}

\subsection{TWB}

As we argue in Sec. \ref{The interferometric scheme}, when TWB state is injected it should be promising to define the observable operator in the form $\widehat{C}\left(\phi_{1},\phi_{2}\right)=\left(N_{1}-N_{2}\right)^{M}, (M>0)$ because of the perfect photon number correlation of TWB. Indeed, in the Sec. \ref{Correlations at the read-out ports} we show that at least up to the second order $M=2$, the strong non-classical correlations are preserved at the output ports of the interferometers (for $\psi=\pi/2$), justifying the conjecture that an advantage in terms of noise reduction would be obtained if we can estimate the phases covariance starting from the measurement of an observable of that form. We notice immediately that for $M=1$, corresponding to the photon numbers subtraction, the proportional coefficient in Eq. (\ref{PSs-Cov}), containing the double derivative with respect to both the phases will be null. Thus, we have to move to the second order measurement i.e.  $\widehat{C}(\phi_{1},\phi_{2})=\left(N_{1}(\phi_{1})-N_{2}(\phi_{2})\right)^{2}= N_{1}^{2}+N_{2}^{2}-2N_{1}N_{2}$. Hereinafter we also consider the same central phase of the two interferometers $\phi_{1,0}=\phi_{2,0}=\phi_{0}$.

According to Eq. (\ref{PSs-Cov}) we get:
\begin{equation}\label{phaseCov-TWB}
  \mathcal{E}_\parallel \left[ \delta \phi_{1} \delta \phi_{2} \right]
  \approx
  \frac{\mathcal{E}_\parallel \left[N_{1}N_{2}
      \right]-\mathcal{E}_\perp \left[N_{1}N_{2}\right]}{\langle \partial_{\phi_{1},\phi_{2}}^{2} N_{1}(\phi_{0})N_{2}(\phi_{0})\rangle },
\end{equation}
where we have used again the symmetry of the statistical properties of the two interferometers, in particular $\mathcal{E}_{\parallel(\perp)}\left[ N_{1}^{2}\right]=\mathcal{E}_{\parallel(\perp)} \left[ N_{2}^{2}\right]$. The covariance of the phase noise is proportional to the difference between the photon number correlation when the phase noise is correlated ($\parallel$) and when it is not ($\perp$), as one could expect.

The uncertainty of the measurement, due to photon noise can be obtained by Eq. (\ref{U0}) where $\mathrm{Var}[ \widehat{C}(\phi_{1,0 }, \phi_{2,0}) ]=\langle\left(N_{1}(\phi_{0}) -N_{2}(\phi_{0}) \right)^{4}\rangle- \langle(N_{1}(\phi_{0}) -N_{2}(\phi_{0}) )^{2}\rangle^{2}$.

\subsection{Independent Squeezed States }
It is rather intuitive that the most simple form of the measurement operator $\widehat{C}(\phi_{1},\phi_{2})$, that combines the squeezed quadratures measured at the read-out port, and has non-null mixed derivative with respect to the phases,  $\partial_{\phi_{1},\phi_{2}}^{2}\widehat{C}\neq0$, would be the product $Y_{1}\cdot Y_{2}$ (where $Y_{i}$ are the squeezed quadratures). However, to avoid the presence of a dc-component in the measurement it turns out more useful to consider the fluctuation of the quadratures around their central value, therefore defining $\widehat{C}=(Y_{1}(\phi_{1})-\mathcal{E} [Y_{1}])\cdot(Y_{2}(\phi_{2})-\mathcal{E} [Y_{2}])$, where we have taken into account that $\mathcal{E}_\parallel  [Y_{i}]=\mathcal{E}_\perp [Y_{i}]=\mathcal{E} [Y_{i}]$. The covariance of the phases is estimated according to Eq. (\ref{PSs-Cov}) as:
\begin{equation}\label{phaseCov-SQ}
  \mathcal{E}_\parallel \left[ \delta \phi_{1} \delta \phi_{2} \right]
  \approx
  \frac{\mathcal{E}_\parallel \left[Y_{1}Y_{2}
      \right]-\mathcal{E}_\perp \left[Y_{1}Y_{2}\right]}{\langle \partial_{\phi_{1},\phi_{2}}^{2} Y_{1}(\phi_{0}) Y_{2}(\phi_{0}) \rangle },
\end{equation}

Since the fluctuations of the quadratures due to quantum noise are independent in the two interferometers, the zero order uncerainty on the measured observable (ses Eq. (\ref{U0})) remains $\mathrm{Var}[ \widehat{C}(\phi_{1,0 }, \phi_{2,0}) ]=\langle\left(Y_{1}(\phi_{0})  -\mathcal{E} [Y_{1}]\right)^{2}\rangle \langle\left(Y_{2}(\phi_{0}) -\mathcal{E} [Y_{2}]\right)^{2}\rangle$.

\section{Results}
\label{Results}
The calculation of the variance of the measurement operator $\widehat{C}(\phi_{0})$, in particular for TWB case, involves many fourth-order terms of the photon number operator, i.e. eighth-order product of field operator $c_{i}$ and $c^{\dag}_{i}$, and the calculation and the complete expression of this variance are too cumbersome to be reported here. Thus, we will present numerical results for the most significative regions inside the parameter space and we give some general expression in particular relevant limits. First of all we need to define the classical benchmark to compare performance using quantum light. The uncertainty achievable in the estimation of the phase covariance, if only the coherent beams are used, is $\mathcal{U}_{\rm CL}^{(0)}=\sqrt{2}/(\eta \mu \cos^{2}\left[\phi_{0}/2\right])$. We notice that the scaling is with the detected number of photons, i.e. the square of the shot noise limit typical of the single phase estimation. This is because we are measuring a second order quantity, namely the covariance of the phases. As usual, it is clear that without any particularly low energy constraint, in order to reach high sensitivity in a phase-correlation measurement is necessary to push the intensity of the classical field. Therefore, even the quantum strategy should face and should improve the sensitivity when high power is circulating into the interferometers. In general we will consider the limit $\mu\gg1$.

Concerning the use of the two independent squeezed states, we can summarize the results in the following couple of equation:
\begin{eqnarray}\label{U-SQ(phi)}
\mathcal{U}_{\rm SQ}^{(0)}/\mathcal{U}_{\rm CL}^{(0)}&\approx& 1-\frac{\eta(1 +  \cos[\phi_{0} ])}{2}+\frac{\eta\cos^{2}[\phi_{0}/2]}{4 \lambda}\quad(\mu\gg1, \lambda\gg1)\\\nonumber
\mathcal{U}_{\rm SQ}^{(0)}/\mathcal{U}_{\rm CL}^{(0)}&\approx& 1-\eta  (1+\cos[\phi_{0} ]) \sqrt{\lambda }(1-\sqrt{\lambda })\quad(\mu\gg1, \lambda\ll1)
\end{eqnarray}

\begin{figure}[htb]
\centering
\includegraphics[width=9cm]{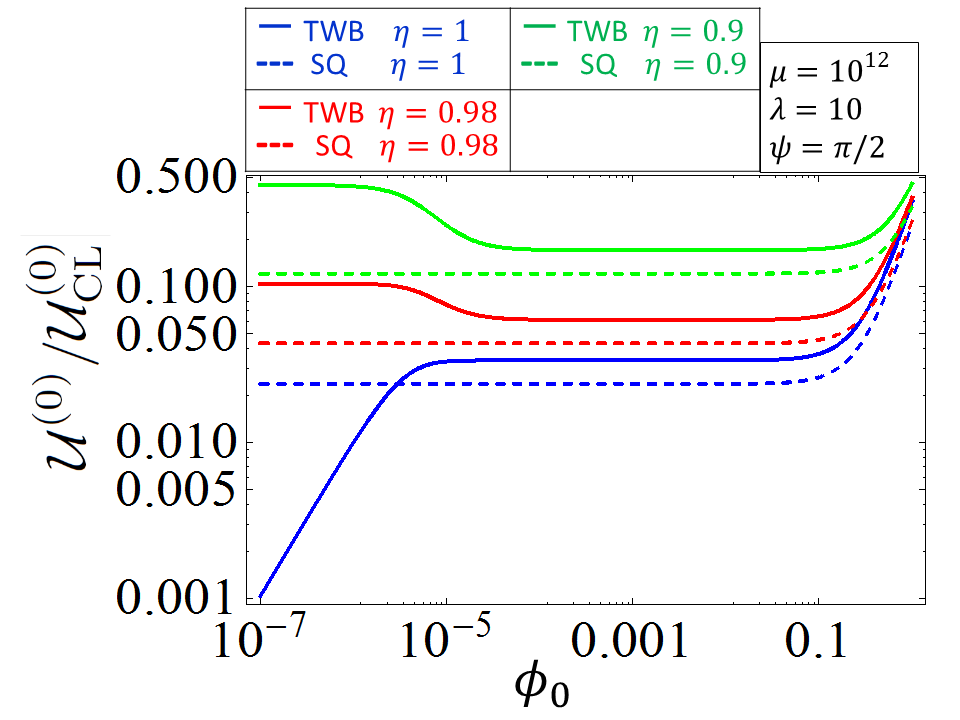} \caption{Uncertainty provided by the twin beam state (TWB) and by the product of squeezed states (SQ) normalized to the classical limit in function of the central phase $\phi_{0}$ in which the interferometers are operated. The plot is obtained for $\lambda=10$, $\mu=3\times10^{12}$, $\psi=\pi/2$, and different colors corresponds to different values of the detection efficiency $\eta$.}
\label{UvsPhi(eta)}
\end{figure}

\begin{figure}[htb]
\centering
\includegraphics[width=9cm]{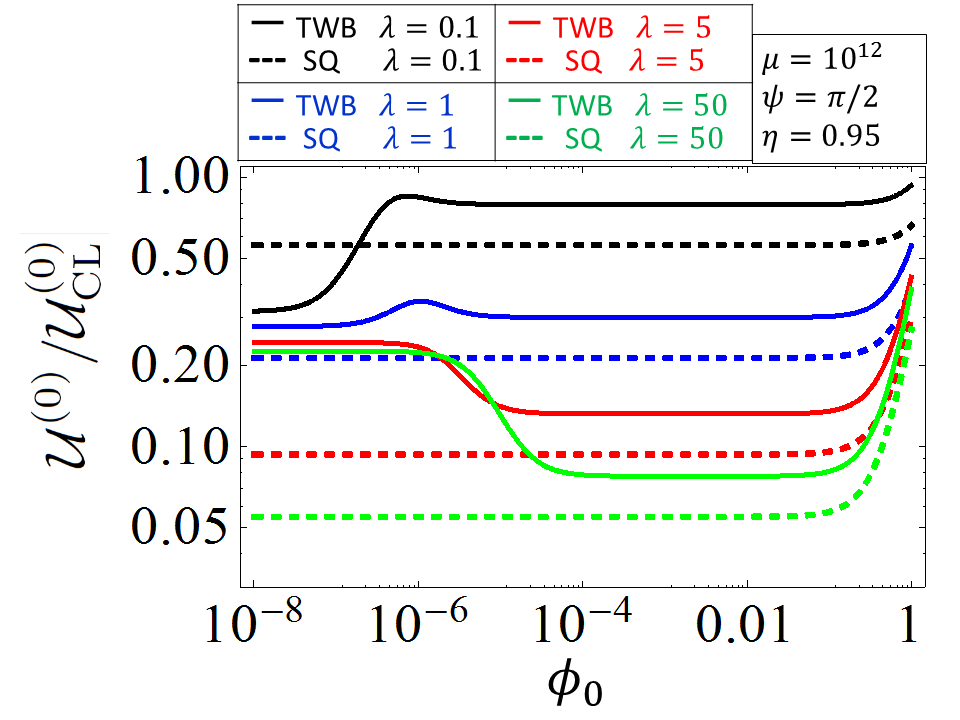} \caption{Uncertainty provided by the twin beam state (TWB) and by the product of squeezed states (SQ) normalized to the classical limit in function of the of the central phase $\phi_{0}$ in which the interferometers are operated. The plot is obtained for $\eta=0.95$, $\mu=3\times10^{12}$, $\psi=\pi/2$, and different colors corresponds to different values of the mean number of photons per mode $\lambda$ of the quantum light.}
\label{UvsPhi(lambda)}
\end{figure}

Of course, we expect that the advantages of using squeezing, and in general quantum light, is effective when it experience a low loss level. Thus the most interesting regime is when the two interferometers transmit almost all the quantum light to the read-out port, meaning that the central phase must be close to 0, according to the BS-like behaviour $\tau=\cos^{2}[\phi_{0}/2]\simeq1$. Applying this limit to Eq.s (\ref{U-SQ(phi)} we have $\mathcal{U}_{\rm SQ}^{(0)}/\mathcal{U}_{\rm CL}^{(0)}\approx 1-\eta+ \eta/4\lambda$ for $\lambda\gg1$ and $\mathcal{U}_{\rm SQ}^{(0)}/\mathcal{U}_{\rm CL}^{(0)}\approx 1-2\eta\sqrt{\lambda}(1-\sqrt{\lambda })$ for $\lambda\ll1$. We can appreciate visually what does it mean by looking to the Fig. \ref{UvsPhi(eta)}. A flat region (in logarithmic scale) appears in the uncertainty reduction in function of the central phase $\phi_{0}$. Since $\lambda=10$ in the figure, the value of the uncertainty reduction, given by the previous expression, is well represented by $1-\eta+ \eta/4\lambda$. In the opposite limit of  $\lambda\ll1$  the advantage of squeezing is lost, according to the second of Eq.s (\ref{U-SQ(phi)}). For example for $\lambda=0.1$, represented in figure Fig. \ref{UvsPhi(lambda)}, the improvement is reduced to 0.6.

Concerning TWB, one can clearly discern two different regions both in Fig. \ref{UvsPhi(eta)} and Fig. \ref{UvsPhi(lambda)}, one for really small values of the central phase, namely $\phi_{0}<10^{-6}$ and an other one in the range $10^{-5}<\phi_{0}<10^{-1}$. They correspond, for the a specific choice of the parameter indicated in figure, to the two relevant regimes that have been individuated in Sec. \ref{Correlations at the read-out ports}.

\begin{figure}[htb]
\centering
\includegraphics[width=11cm]{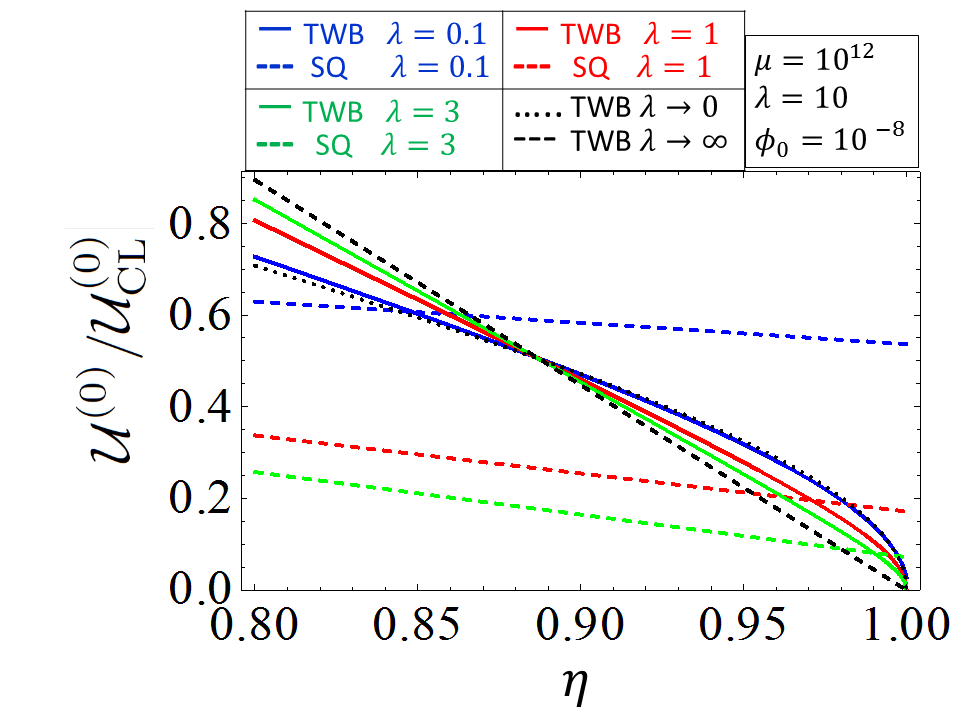} \caption{Uncertainty provided by the twin beam state (TWB) and by the product of squeezed states (SQ) normalized to the classical limit in function of the quantum efficiency $\eta$. The plot is obtained for $\phi=10^{-8}$, $\mu=3\times10^{12}$, $\psi=\pi/2$, and different colors correspond to different values of the mean number of photons per mode $\lambda$ of the quantum light.}
\label{UvsEta}
\end{figure}
\begin{description}
  \item[(A)-TWB like correlations-] when $\langle N\rangle^{coh}_{\eta\tau}\ll \langle N\rangle^{TWB}_{\eta\tau}$. The condition is guaranteed if the central phases are close enough to zero, $\phi_{1,0}=\phi_{2,0}\simeq0 $, meaning that the transmissivity of the equivalent-BS approaches the unity $\tau_{i}=Cos^{2}[\phi_{i,0}/2]\simeq1 (i=1,2)$. This is the regime studied and reported in \cite{prl}. For intense coherent beam and intense TWB source, i.e. $\mu\gg\lambda\gg1$,  one gets $\mathcal{U}_{\rm TWB}^{(0)}/\mathcal{U}_{\rm CL}^{(0)} \approx 2\sqrt{5}\left( 1-\eta\right)$, while in the case of faint TWB, $\lambda\ll1$ and $\mu\gg1$, the result is  $\mathcal{U}_{\rm TWB}^{(0)}/\mathcal{U}_{\rm CL}^{(0)} \approx \sqrt{2(1-\eta)/\eta}$. In both cases TWB allows reaching an amazing uncertainty reduction that approaches zero for unitary detection efficiency. This behaviour is clearly shown in Fig. \ref{UvsEta} in which the choice of $\mu$ and $\phi_{0}$ ensure to be in the TWB-like regime at least for the range of values of $\lambda$ represented there. All the TWB curves drop to zero, and for some value of the efficiency, depending on the intensity $\lambda$, they fall below the corresponding SQ curves. The limits for $\lambda\ll1$ and for $\lambda\gg1$ are also reported in dotted and dashed black lines respectively. However, we observe that for quantum light intensity $\lambda>1$ reachable in experiments nowadays (for example $\lambda=3$ in the picture) squeezing performs far better than TWB except for extremely demanding overall detection efficiency.

  \item[(B)-Bright quantum correlations-] when $\langle N\rangle^{coh}_{\eta\tau}\gg \langle N\rangle^{TWB}_{\eta\tau}$. This regime corresponds to the flat region shown in Fig.s \ref{UvsPhi(eta)} and \ref{UvsPhi(lambda)} for intermediate values of the central phase $\phi_{0}$. Aside a constant factor, the uncertainty reduction for $\mu\gg1$ behaves as for the two independent squeezing case, specifically $\mathcal{U}_{\rm TWB}^{(0)}/\mathcal{U}_{\rm CL}^{(0)}=\sqrt{2}\mathcal{U}_{\rm SQ}^{(0)}/\mathcal{U}_{\rm CL}^{(0)}$. It can be easily appreciated  when comparing the corresponding curves for TWB and Squeezing in the figures (taking in to account the logarithmic scale).
\end{description}

For the sake of completeness, we did the same analysis considering to exploit the anti-correlations, defining the observable as  $\widehat{C}(\phi_{1},\phi_{2})=\left(N_{1}+N_{2}\right)^{2}= N_{1}^{2}+N_{2}^{2}+2N_{1}N_{2}$ (for $\psi=0$) instead of the correlation when TWB are injected, obtaining analogous results in the regime of $\mu\left(1-\tau\right) >> \lambda\tau$.

\section{Discussion and conclusions}
\label{conclusions}

In Sec. \ref{Results}  we observed interesting features, leading to promising experimental conditions. Referring to Fig.  \ref{UvsPhi(eta)} and \ref{UvsPhi(lambda)} there is an extended range of value of the central working phase $\phi_{0}$ of the interferometers in which the uncertainty reduction achievable by adopting quantum light is stable, at the value $\mathcal{U}_{\rm }^{(0)}/\mathcal{U}_{\rm CL}^{(0)}\approx 1-\eta+ \eta/4\lambda$ both with SQB and TWB (a part a factor $\sqrt{2}$ in the last case). This kind of scaling is a well known results of phase estimation in a single interferometer combining coherent strong field and single mode squeezed light (in fact  $4\lambda\approx e^{2|\xi|}$ in the limit $4\lambda\gg1$) \cite{cav}. Therefore, it turns out that a measurement of the phase correlation retains the same advantage of the single phase estimation. As an example, for $\eta=0.9$  and $\lambda=3$, compatible with the actual technology, we have 5.7 times of uncertainty reduction in the single measurement. Since in any experiments $\mathcal{N}$ measurements are performed and the final uncertainty is $\mathcal{U}/\sqrt{\mathcal{N}}$, one would easily obtain the same sensitivity with a number of runs 30 times smaller, hence reducing the total measurement time of the same amount.

On the other side, only for TWB, there exists a special setting of the central phase of the interferometer, when the classical fields component is made negligible at the read-out ports with respect to the TWB component, in which in principle the uncertainty reach the zero point, whatever the intensity of TWB. In particular for faint TWB ($\lambda\ll1$) the uncertainty scales as $\mathcal{U}_{\rm TWB}^{(0)}/\mathcal{U}_{\rm CL}^{(0)} \approx \sqrt{2(1-\eta)/\eta}$ while for intense TWB ($\lambda\gg1$) one gets $\mathcal{U}_{\rm TWB}^{(0)}/\mathcal{U}_{\rm CL}^{(0)} \approx 2\sqrt{5}\left( 1-\eta\right)$. Even if at first glance this looks rather exciting, Fig. \ref{UvsEta} shows that in terms of absolute sensitivity squeezing performance can be overtaken only for rather high detection efficiency. For example for $\lambda=3$ we expect $\mathcal{U}_{\rm TWB}^{(0)}<\mathcal{U}_{\rm SQ}^{(0)}$ for $\eta\geq0.99$. However, for limited quantum resources, namely $\lambda<1$, TWB performs better than squeezing already for smaller and more realistic efficiency values.

In conclusion, we have analyzed in detail a system of two interferometers aimed at the detection of extremely faint phase-fluctuation. The idea behind is that a correlated phase-signal like the one introduced by the ``holographic noise'' could emerge by correlating the output ports of the interferometers, even when in the single interferometer it confounds with the background. We demonstrated that injecting quantum light in the free ports of the interferometers can reduce the photon noise of the system beyond the shot-noise, enhancing the resolution in the phase-correlation estimation. Our results basically confirms the benefit of using squeezed beams together with strong coherent beams in interferometry, even in this correlated case. However, mainly we concentrated on the possible use of TWB, discovering interesting and probably unexplored areas of application of bipartite entanglement and in particular the possibility of reaching in principle surprising uncertainty reduction.

{\bf Acknowledgements}

This work has been supported by EU (BRISQ project) and MIUR (FIRB ``LiCHIS'' -- RBFR10YQ3H). This publication was made possible through the support of a grant from the John
Templeton Foundation. The opinions expressed in this publication are those of the
authors and do not necessarily reflect the views of the John Templeton Foundation.

\end{document}